\documentclass[twocolumn,prb,aps,10pt]{revtex4-1}
\usepackage{graphicx,epsf}

\begin{document}

\title{Single-parameter adiabatic charge pumping in carbon nanotube resonators}

\author{C.A. Perroni$^{1,2}$, A. Nocera$^{3}$, and V. Cataudella$^{1,2}$}

\affiliation{$^{1}$CNR-SPIN and $^{2}$Universita' degli Studi di Napoli Federico
II, Complesso Universitario Monte Sant'Angelo, Via Cintia, I-80126
Napoli, Italy \\
$^{3}$Department of Physics, Northeastern University, Boston, Massachusetts 02115, USA}

\begin{abstract}
Single-parameter adiabatic charge pumping, induced by a nearby radio-frequency antenna, is achieved in suspended carbon nanotubes close to the mechanical resonance. The charge pumping is due to an important dynamic adjustment of the oscillating motion to the antenna signal and it is different from the mechanism active in the two-parameter pumping. Finally, the second harmonic oscillator response shows an interesting relationship with the first harmonic that should be experimentally observed.
\end{abstract}

\maketitle
\newpage

\section{Introduction}

An electronic current flowing into a conductor normally requires a bias voltage $eV_{bias}$ (where
$e$ is the electron charge) applied between the electrodes. However, even at zero bias,
ac fields are able to induce a finite current through nanoscopic systems where the electron
coherence length is larger than the device size. This phenomenon is known as charge pumping. \cite{thouless,altshuler}
This effect has been mostly realized in the adiabatic regime (driving frequency $\omega$ much smaller
than $\Gamma$, the inverse of time spent by an electron to cross the sample) with two out of phase driving parameters, for example
left and right lead voltages or one lead and the gate voltages. \cite{brouwer} Adiabatic charge pumping has been realized in
carbon-based devices such as carbon nanotube dots. \cite{buite,wei}

Recent experiments have shown that pumping with a single-parameter is also feasible. \cite{kaest,fuji,kaest1}
Single-parameter pumping represents an important step forward for different reasons. First,
devices with one single driving source can be more easily fabricated. Then, one parameter involves a reduction of device
size and dissipation. Although single-parameter pumping has been realized in carbon-based devices, they exhibit an important drawback: higher frequencies are normally needed to get currents similar to those of two-parameter pumping. \cite{vavi,mosk,torres,agar,torres1}
Therefore, the quest for single-parameter pumping in the adiabatic regime represents a relevant issue.

Recent experiments \cite{Steele,Huttel} have characterized carbon nanotube devices working in the
adiabatic regime (driving external frequency $\omega$ in MHz range, $\Gamma$ of the order of tens of GHz)
with an extremely large quality factor $Q$ ($Q>10^{5}$). By measuring the variations of the electronic current flowing through the
nanotube as a function of frequency of a nearby antenna actuating its motion, very well defined resonances corresponding to the
bending mode of the nanotube itself can be detected. \cite{Huttel1,meerw} Moreover, by adjusting the antenna power, the nanotube resonator can easily be
tuned into the nonlinear vibration regime. However, the bias voltages applied in the experiment are large, typically $eV_{bias}>>k_{B}T$,
with $k_{B}$ Boltzmann constant and temperature $T$ of the order of mK. \cite{Steele}   Only recently, in the same set-up, measurements
have been focused under zero bias. \cite{ganzhorn} However, the actuation power has been kept very low analyzing only the linear regime of the oscillator and charge conductance.

In this work, we propose to use the existing set-up for suspended carbon nanotubes in order to investigate the zero bias regime with increasing the antenna power. We find theoretical evidence that the antenna is able to pump sufficient charge close to the mechanical resonance making single-parameter adiabatic charge pumping feasible in carbon nanotube resonators. The pumping mechanism is based on an important dynamic adjustment of the mechanical motion of the nanotube to the external drive. The new pumping process in the weakly non-linear regime for the oscillator differs from that in the two-parameter pumping. Actually, it does not rely on a phase shift, but it involves relevant differences of the current response already on a single period of the antenna signal. Finally, an interesting prediction is made about the second harmonic oscillator response and its relationship with the first harmonic.

In the small frequency window of interest for the resonant behavior, the electronic part of the device is modeled in terms of a few electronic
levels coupled to the leads through standard tunneling terms. \cite{Steele,Huttel,Blanter,Blanter1,Bennett,Weick,genovesi}
The interaction between the charge on the dot and the oscillating degrees of freedom  has been shown \cite{Nocera,alberto,alberto1} to be equivalent to a Holstein-like coupling. \cite{Hol} The effects induced by the external antenna give an assigned time dependence to the gate voltage.
Therefore, the overall Hamiltonian for a simplified model is
\begin{equation}\label{Htot}
\hat{\cal H}(t)=\hat{\cal H}_{el}(t)+ \hat H_{osc} +{\hat H}_{int}. \label{htot}
\end{equation}

The electronic Hamiltonian $\hat{\cal H}_{el}(t)$ is given by
\begin{equation}\label{Hel}
\hat{\cal H}_{el}(t)=
\sum_{k,\alpha}\varepsilon_{k,\alpha}{\hat c^{\dag}_{k,\alpha}}{\hat c_{k,\alpha}}+
\sum_{k,\alpha}(V_{k,\alpha}{\hat c^{\dag}_{k,\alpha}}{\hat d}+
h.c.)+E_{G}(t) {\hat N_{el}}, \label{hele}
\end{equation}
where the operators ${\hat c^{\dag}_{k,\alpha}} ({\hat
c}_{k,\alpha})$ create (annihilate) electrons with momentum $k$
and energy $\varepsilon_{k,\alpha}=E_{k,\alpha}-\mu_{\alpha}$ in
the left ($\alpha=L$) or right ($\alpha=R$) free metallic leads,
while the electronic tunneling between the nanotube level and a
state in the lead has amplitude $V_{k,\alpha}$. The
nanotube's electronic level has time-dependent energy $E_{G}(t)$ due to
the antenna, ${\hat d^{\dag}} ({\hat d})$ are creation (annihilation) operators, and
${\hat N_{el}}=\hat d^{\dag} \hat{d}$ is the electronic occupation operator on the nanotube. \cite{Nota1}
The level energy is $E_{G}(t)=e V_{ext} cos(\omega_{ext})+e V_G$,
where $V_{ext}$ is the amplitude of the external antenna potential, $\omega_{ext}$
is the driving frequency, $V_G$ is the static gate potential.
The chemical potentials in the leads $\mu_{L}$ and $\mu_{R}$ are assumed to be
equal to zero. The coupling to the leads is described by the tunneling rate
$\Gamma_{\alpha,k}=2\pi\rho_{\alpha}|V_{k,\alpha}|^{2}/\hbar$,
where $\rho_{\alpha}$ is the density of states in the lead
$\alpha$. We will suppose symmetric coupling
($\Gamma_{L,k}=\Gamma_{R,k}$) and a flat density of states for the
leads, considered as thermal baths at temperature $T$, within the
wide-band approximation: $\Gamma_{\alpha,k}\mapsto
\Gamma_{\alpha}$, with $\alpha=L,R$, and $\Gamma= \Gamma_{L}+\Gamma_{R}$.

In Eq. \ref{htot}, the Hamiltonian of the mechanical degree of freedom is
$\hat H_{osc}={\hat p^{2}\over 2m} + {1\over 2}m
\omega^{2}_{0}\hat x^{2}$, with $\hat x$ and $\hat p$ the position and momentum operator, respectively, $\omega_{0}$ the frequency
and $m$ the mass ($k=m\omega_{0}^{2}$ elastic constant). The electron-oscillator
interaction in Eq. \ref{htot} is ${\hat H}_{int}=\lambda \hat x {\hat
N_{el}}$, where $\lambda$ is the coupling strength.
We point out that the main linear and non-linear features observed in suspended carbon nanotubes \cite{Steele,Huttel} have been reproduced
considering a single vibrational mode in the 120-300 MHz range suggesting the validity
of the adiabatic limit ($\omega_{0}/\Gamma <<1$). \cite{alberto}  Moreover, estimates of
the effective electron-oscillator coupling $E_{p}$ ($E_{p}={\lambda^2\over 2k}\simeq5\mu e V$) indicate a strong
interaction ($E_{p}/\hbar\omega_{0} \simeq 10$). \cite{Weick,alberto}

The regime of the parameters relevant for experimental results, but also for the effects
discussed in this work is: $\hbar\omega_{0}<<E_{p} \sim k_{B}T << \hbar\Gamma$.
Theoretical treatments of dots coupled to oscillators
including external drives have been recently considered. \cite{Labadze,Gorelik,Armour}
In these works, tunneling is considered in the sequential regime, therefore quantum effects in
the electronic dynamics such as cotunneling are disregarded.
We point out that these effects are important since $k_{B}T$ is much smaller than
$\hbar\Gamma$. Our approach, based on non equilibrium Green functions, \cite{haug} takes
into account from the beginning all higher order tunneling terms. Moreover, as shown in the supplementary material,
the temperature is an important parameter controlling the amount of pumped charge. Finally, when $k_{B}T >> \hbar\omega_{0}$, the semi-classical treatment of
the oscillator dynamics is well justified in organic materials. \cite{alberto,Millis,perroni1,perroni2,perroni3}

The adiabatic approach can be applied to the electronic dynamics due to slow temporal perturbations:
$\omega_0 << \Gamma$, $ \dot{E}_{G}(t) << \hbar \Gamma^2$, with  $\dot{E}_{G}(t)$ time derivative of  $E_{G}(t)$. \cite{citro}
Therefore, the electronic level can be described by an effective energy
$E_{G}^{eff}(x,t)=E_{G}(t)+\lambda x$, with $ \dot{E}^{eff}_{G}(v,t)=\dot{E}_{G}(t)+\lambda v$.
In particular, the adiabatic expansion of the dot occupation $N_{el}(x,v,t)$ gives
\begin{equation}
N_{el}(x,v,t) \simeq N_{el}^{(0)}(x,t)+N_{el}^{(1)}(x,v,t),
\end{equation}
where the zero order "static" term $N_{el}^{(0)}(x,t)$ is
\begin{equation}
N_{el}^{(0)}(x,t)= \int_{- \infty}^{+ \infty}
{d (\hbar \omega) \over 2\pi}  \frac{ f(\hbar \omega) \hbar  \Gamma }{ \left[ \hbar \omega - E_{G}^{eff}(x,t) \right]^2+\frac{ (\hbar \Gamma)^2 }{4}},\label{enne0}
\end{equation}
dependent on the Fermi distribution of the leads $f(\hbar \omega)$, and the first order "dynamic" term $N_{el}^{(1)}(x,v,t)$ is
\begin{equation}
N_{el}^{(1)}(x,v,t)= \frac{\hbar}{2}
\dot{E}^{eff}_{G}(v,t) (\hbar \Gamma) R(x,t),\label{enne1}
\end{equation}
with
\begin{equation}
R(x,t) =
\int_{- \infty}^{+ \infty}
{d (\hbar \omega) \over 2\pi}
\frac{  g(\hbar \omega) \hbar \Gamma}
{ \left( [ \hbar \omega - E_{G}^{eff}(x,t)]^2+\frac{[\hbar \Gamma]^2}{4} \right)^2},\label{errefun}
\end{equation}
dependent on the derivative $g(\hbar \omega)=- \partial f(\hbar \omega) / \partial (\hbar \omega)$.

Central for this work is the adiabatic expansion of the electronic current $J_{\alpha}(x,v,t)$ from the lead $\alpha$ to the dot.
We stress that the zero order vanishes due to absence of bias voltage:
$J_{\alpha}(x,v,t) \simeq J_{\alpha}^{(1)}(x,v,t)$, where
\begin{equation}
J_{\alpha}^{(1)}(x,v,t) = -e  \dot{E}_{G}^{eff}(v,t) (\hbar \Gamma_{\alpha}) V(x,t), \label{curr}
\end{equation}
with
\begin{equation}
V(x,t) = \int_{- \infty}^{+ \infty}
{d (\hbar \omega) \over 2\pi}  \frac{ g(\hbar \omega)}
{ [ \hbar \omega -E_{G}^{eff}(x,t)]^2+\frac{(\hbar \Gamma)^2}{4} }.
\end{equation}
is proportional to the time derivative of the effective level energy. We will see that the self-consistent action of the oscillator makes zero the current contribution directly due to its velocity. Moreover, charge conservation is valid: $e \dot{N}_{el}^{(0)}(x,t)=J_L^{(1)}(x,v,t)+J_R^{(1)}(x,v,t)$.

The back-action of the nanotube motion on the current is of paramount importance.
Within the adiabatic approach, \cite{Nocera} the oscillating dynamics of
the nanotube can be described by a nonstandard Langevin
equation which takes into account the interaction with both the nanotube itself and the
electronic environment given by the macroscopic leads. This equation is controlled by a self-consistent effective
anharmonic force as well as by damping and fluctuating terms which, due to the presence of the antenna, depend not only
on the resonator displacement $x$, but also explicitly on time $t$.
This complex oscillator dynamics represents one of the main results of this work since it allows to explore
in a self-consistent dynamic way the non-linear non-perturbative response.

The force on the oscillator can be decomposed into two adiabatic expansion terms:
\begin{equation}\label{for}
F(x,v,t)=-k x - \lambda N_{el}(x,v,t) \simeq F^{(0)}(x,t)+F^{(1)}(x,v,t). \label{force1}
\end{equation}
The zero order force $F^{(0)}(x,t)$ represents the "static" part sensitive to the average charge occupation: $F^{(0)}(x,t)=-kx -\lambda N_{el}^{(0)}(x,t)$,
with $N_{el}^{(0)}(x,t)$ zero order electron density given in Eq. (\ref{enne0}). Instead, the first order "dynamic" force $F^{(1)}(x,v,t)$ is sensitive to charge fluctuations. Moreover, it contains not only a dissipative term proportional to the velocity, but also a very complex non-linear term due to the effects of the external antenna:
\begin{equation}
F^{(1)}(x,v,t)=-\lambda N_{el}^{(1)}(x,v,t)=-A(x,t) v + B(x,t) \dot{E}_{G}(t),
\end{equation}
with $N_{el}^{(1)}(x,v,t)$ given in Eq. (\ref{enne1}), the damping coefficient $A(x,t)= \frac{\hbar \lambda^2}{2}  (\hbar \Gamma) R(x,t)$
positive definite, the term $B(x,t)= - \frac{\hbar \lambda}{2} (\hbar \Gamma)  R(x,t)$,
and $R(x,t)$ given in Eq. (\ref{errefun}).

Within the adiabatic approach, one derives the following fluctuating oscillator force term \cite{alberto} due the
"fast" electronic charge motion within the nanotube:
\begin{equation}\label{fluforce}
\langle F(x,t) F(x,t')\rangle=
D(x,t) \delta(t-t').
\end{equation}
Due to zero voltage bias, the fluctuation-dissipation condition is verified at each fixed position x and time t:
$D(x,t)=2 k_B T A(x,t)$. The oscillator can be locally equilibrated in space and time, so that the velocity distribution function is of Boltzmann type. Therefore, in Eq.(\ref{curr}), the term proportional to the oscillator velocity vanishes and only the current term directly due to the antenna gives a net contribution.

The resulting complex Langevin equation with time-dependent coefficients and multiplicative noise is:
\begin{eqnarray}\label{Langevin1}
m \dot{v} &=& - A(x,t) v+F_{det}(x,t)+ \sqrt{D(x,t)}\xi(t),\\
\langle\xi(t)\rangle&=&0,\;\;\;\;\langle\xi(t)\xi(t')\rangle=\delta(t-t'),\nonumber
\end{eqnarray}
where $F_{det}(x,t)=F^{(0)}(x,t)+B(x,t) \dot{E_G}(t)$ is the deterministic part of the force,
and $\xi(t)$ is a standard white noise term. Since the external antenna term is periodic over $T_{ext}$, the coefficient $A(x,t)$ and the force $F_{det}(x,t)$
are periodic as well. Exploiting this periodicity, we numerically solve the Langevin equation ({\ref{Langevin1}})
calculating the oscillator distribution function $P(x,v,t)$ and reduced position distribution function $P(x,t)$.

We average an electronic or an oscillator observable $O(x,v,t)$:
$<O>(t)=\int d x \int d v P(x,v,t) O(x,v,t)$.
Given the periodicity over $T_{ext}$, the time average $\bar{O}$ is $\bar{O}=\frac{1}{T_{ext}} \int_{0}^{T_{ext}} d t <O>(t)$.
Due to charge conservation, $\bar{J}_R=-\bar{J}_L$, so that the central quantity of this work is $Q=T_{ext}(\bar{J}_L-\bar{J}_R)/2$, the charge pumped in a period.

\begin{figure}
\centering
{\includegraphics[width=8.5cm,height=9.5cm,angle=0]{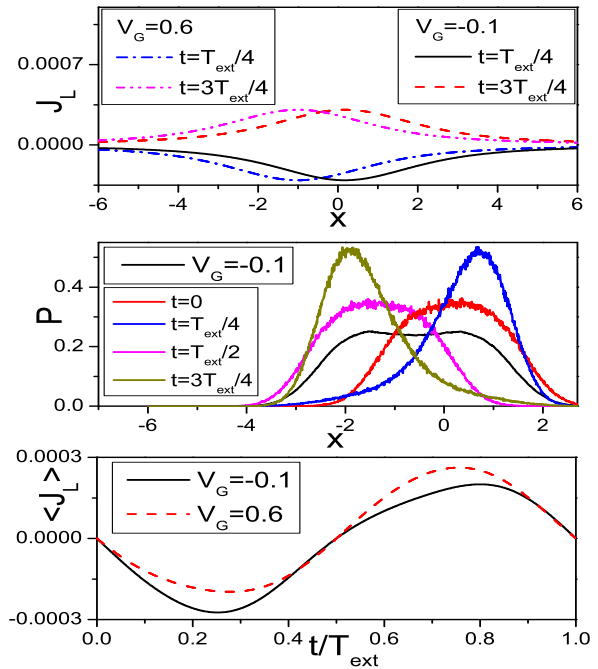}}
\caption{(Color online) Upper Panel: Left current $J_L$ as a function of oscillator
position x for different times t and static gate voltages $V_G$. Middle Panel: Reduced position probability
distribution P as a function of x for different times t at fixed static voltage $V_G$ (the time averaged distribution is in black line). Lower Panel:
The averaged value $<J_L>$ as a function of time t for different gate voltages $V_G$. In this plot, $V_{ext}=0.1$, $T=0.3$, $E_p=0.3$, and $\omega_{ext}=0.93 \omega_0$, with $T_{ext}=2 \pi / \omega_{ext}$.} \label{fig1}
\end{figure}

In this work, we choose the ratio $\omega_{0}/\Gamma=0.05$ realistic for experimental set-ups. We will measure lengths in units of $x_{0}=\lambda / k$, frequencies in terms of $\Gamma$, energies in units of $\hbar\Gamma$, and we assume $-e=1$, $\hbar=1$, $k_B=1$.

At first, we have analyzed the mechanism through which charge can be pumped in a period into the device in conditions of resonance between the antenna
and the resonator. As discussed in the final part of this work, the resonance condition is met for $\omega_{ext}$ smaller than $\omega_0$, since the dot occupation induces a strong softening of the bare frequency. \cite{alberto1} As reported in the lower panel of Fig.\ref{fig1} (where $\omega_{ext}=0.93 \omega_0$), the current $<J_L>$ in the first half-period shows a different behavior from that in the second half-period. This involves that the average on a period is different from zero, allowing to pump charge into the nanotube. In the supplementary material, we stress that the differences in the two half-periods are enhanced close to the resonance.

In order to explain the new mechanism proposed in this work, in the upper panel of Fig.\ref{fig1}, we analyze the left current $J_L(x,t)$ at a quarter and tree quarter of the period $T_{ext}$ for different values of static gate $V_G$ at the resonance. We consider values of gates quite symmetric to electron-oscillator coupling $E_p$ ($E_p=0.3$ in the figure). Actually, there is a renormalization of the dot level due to the coupling with the oscillator, so that, for $V_G=E_p$, the dot is half full. At fixed $V_G$, $J_L(x,t)$ acquires a minimum at a quarter of period, while a maximum at tree quarter of period. Moreover, the
static gate induces a shift of the curves toward positive values for negative $V_G$, but negative values for positive $V_G$.

The shifts of $J_L$ are compared with the behavior of $P(x,t)$ at the resonance. As shown in the middle panel of  Fig.\ref{fig1}, in addition to the new center of the distribution due to the coupling $E_P$, the distribution averaged on a period (black line) not only shows large deviations, but it is bimodal due to the resonance phenomenon. In the supplementary material, we point out that the bimodal character is present only close to the resonance. In the same panel, we have analyzed the behavior of the distribution for different times finding that the two peaks of the averaged distribution are mainly due to the contributions coming from one quarter and tree quarter of a period. At these times, the distribution is largely asymmetric around the maxima of the orbit of the oscillator. Therefore, the tail of the distribution is always able to intercept a spatial region where $J_L$ is not zero. In any case, the probability weights at one quarter and three quarter of the period are different and will affect $<J_L>$ shown in the lower panel. This effect is at source of the pumping mechanism which is due to a relevant dynamical adjustment of the oscillator to the external drive. We point out that this behavior cannot be understood in terms of a phase shift between the external drives, such as in the two-pumping parameter mechanism, \cite{brouwer} or between the ac gate voltage and the parametrically excited mechanical oscillations. \cite{Gorelik}
\begin{figure}
\centering
{\includegraphics[width=9.0cm,height=8.0cm,angle=0]{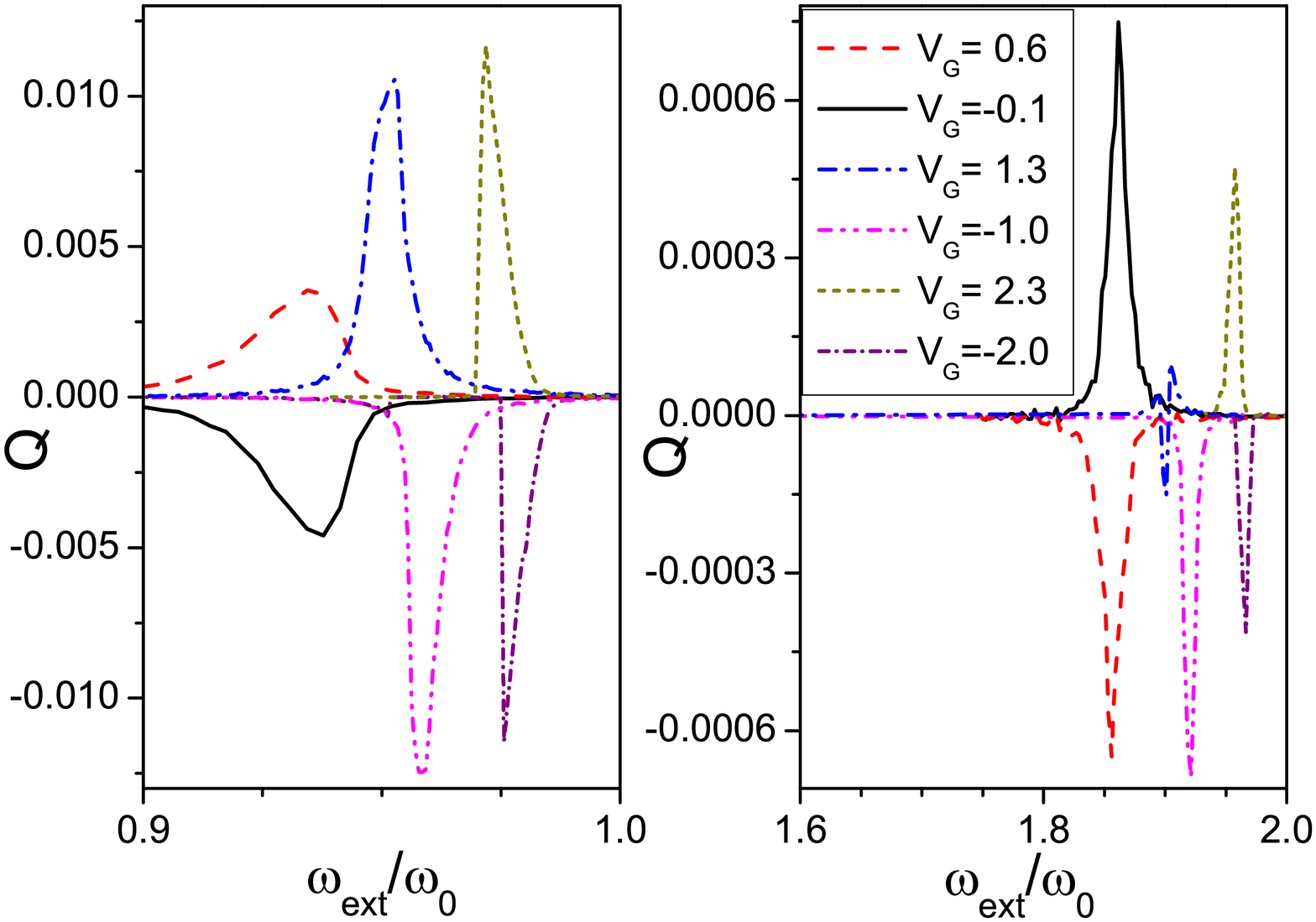}}
\caption{(Color online) Left Panel: Charge Q (in units of e) as a function of the external frequency in the
interval close to $1$ $\omega_0$ with varying the static gate $V_G$. Right Panel: Charge Q (in units of e) as a function of the external frequency in the
interval close to $2$ $\omega_0$ with varying the static gate $V_G$. In this plot, $V_{ext}=0.1$, $T=0.3$, and $E_p=0.3$.}\label{fig2}
\end{figure}

Next, we have investigated the dependence of the pumped charge on the external frequency for different values of the static gate potential $V_G$. As shown in the supplementary material, in the linear regime ($V_{ext}=0.01$), the pumped charge is small and the frequency response is Lorentzian, in agreement with recent experimental results. \cite{ganzhorn} As shown in the left panel of Fig. \ref{fig2}, in the weakly non-linear regime ($V_{ext}=0.1$), the order of magnitude of the pumped charge increases, and it becomes about $0.01 e$. Assuming $\Gamma \simeq 50 GHz$, the pumped current is of the order of $ e \Gamma \leq 1 pA$ at $T\simeq 10 mK$, a value that is susceptible to measurements. This value can be strongly enhanced if  the antenna power increases or the temperature decreases (see supplementary material).

We notice that the sign of the pump depends on that of $V_G$. As reported in the lower panel of Fig.\ref{fig1}, this is due due to the different behavior of the currents for positive and negative $V_G$. As a result, there is a specular symmetry with respect to $V_G-E_p$ ($E_p=0.3$ in Fig. \ref{fig2}). Moreover, with increasing $V_G$, a different regime of response is clearly visible. Actually, the shape tends to be more triangular as a function of the frequency, meaning that the response becomes progressively non-linear with features of the Duffing oscillator. \cite{alberto,Nayfeh} In the supplementary material, we have investigated the fully non-linear regime ($V_{ext}>0.1$)  showing that the first harmonic response shows a saturation with increasing the antenna power. Actually, part of the response is transferred to higher harmonics with the possibility to change the sign of the pumped charge.

\begin{figure}
\centering
{\includegraphics[width=8.5cm,height=9.5cm,angle=0]{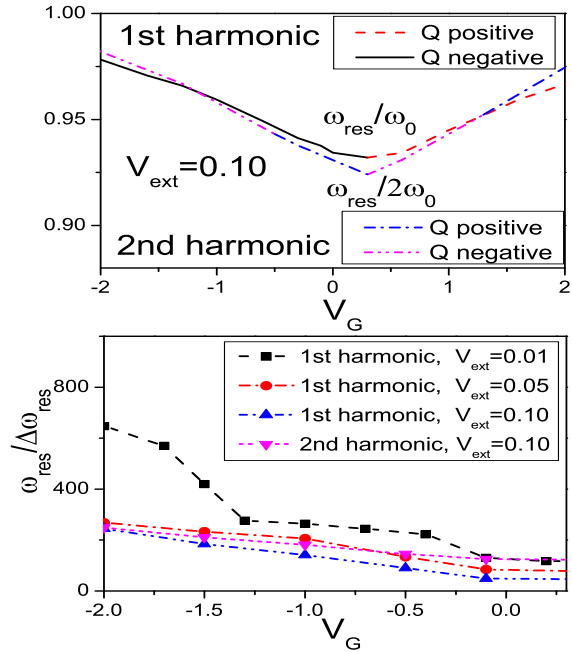}}
\caption{(Color online) Upper Panel: Softening of the resonance frequency corresponding
to the first and to the second harmonic as a function of the static gate voltage $V_G$. Lower Panel: $\omega_{res}/\Delta \omega_{res}$ as a function
of the static gate voltage $V_G$ for different values of the external amplitude $V_{ext}$. In this plot, $T=0.3$, and $E_p=0.3$.}\label{fig3}
\end{figure}

The response for frequencies close to $2 \omega_0$ is very interesting. As shown in the right panel of Fig. \ref{fig2}, in the weakly non-linear regime ($V_{ext}=0.1$), some charge (less than ten per cent of that corresponding to the first harmonic) is pumped close to those frequencies. Moreover, the frequency response shows a very complex behavior with several maxima and minima. Therefore, we have analyzed the position of the frequency peaks of the pumped charge, that is the resonance frequency $\omega_{res}$, and $\Delta \omega_{res}$, the width of the peaks at half height. In the linear regime, the ratio $\omega_{res} / \Delta \omega_{res}$ provides the quality factor of the oscillator. As shown in the upper panel of Fig. \ref{fig3} ($V_{ext}=0.1$, slightly non-linear regime), the first harmonic resonance acquires a characteristic softening which reduces departing from half-filling. We point out that the softening is symmetrical with respect to $V_G-E_P$ even if the pumped charges have opposite signs. Quite surprisingly, when compared to $2 \omega_0$, the second harmonic resonance shows a behavior similar to the first harmonic. This points towards a similarity in the softening of different harmonics and it should be accessible to the experimental confirmation with increasing the antenna power.

Finally, in the lower panel of Fig. \ref{fig3}, we report that the ratio $\omega_{res} / \Delta \omega_{res}$ gets larger with increasing $V_{ext}$. Moreover, it is larger for values of $V_G$ where the nanotube is almost full or empty. Therefore, there is a strong correlation between the shape (Lorentzian/triangular) of the response and its width. Of course, in the slightly non-linear regime, the second harmonic only gives a minor contribution to the response.

In conclusion, for frequencies close to the mechanical resonance, an external antenna is able to realize single-parameter adiabatic charge pumping. The new mechanism is different from that active in the two-parameter pumping since it requires a dynamic adjustment of the mechanical motion of the nanotube to the external drive. Finally, the excitation of the second harmonic is feasible showing a similarity of the softening with the first harmonic. We believe that the main conclusions will not be qualitatively modified by the inclusion of more realistic nanotube levels and interactions \cite{roche} since the results discussed in this work rely on general features of electron and oscillator response.


\begin{thebibliography}{99}
\bibitem{Nota1} We consider for simplicity spin-less electrons, as the inclusion
of spin as well as on-site Coulomb repulsion doesn't qualitatively
change our results.
\end{thebibliography}
\end{document}